\documentclass[%
 reprint,
 superscriptaddress,
 amsmath,amssymb,
 aps,
]{revtex4-2}

\usepackage{graphicx}
\usepackage{dcolumn}
\usepackage{bm}
\usepackage{multirow}
\usepackage{color,soul}

\begin{document}


\title{Topological invariant and domain connectivity in moiré materials}

\author{Ikuma Tateishi}  \email{ikuma.tateishi@riken.jp}
 \affiliation{RIKEN Center for Emergent Matter Science, Wako, Saitama 351-0198, Japan}
\author{Motoaki Hirayama} \email{hirayama@ap.t.u-tokyo.ac.jp}
 \affiliation{Department of Applied Physics, University of Tokyo, Tokyo 113-8656, Japan}
 \affiliation{RIKEN Center for Emergent Matter Science, Wako, Saitama 351-0198, Japan}

\date{\today}

\begin{abstract}
Recently, a moiré material has been proposed in which multiple domains of different topological phases appear in the moiré unit cell due to a large moiré modulation. Topological properties of such moiré materials may differ from that of the original untwisted layered material. In this paper, we study how the topological properties are determined in moiré materials with multiple topological domains. We show a correspondence between the topological invariant of moiré materials at the Fermi level and the topology of the domain structure in real space. We also find a bulk-edge correspondence that is compatible with a continuous change of the truncation condition, which is specific to moiré materials. We demonstrate these correspondences in the twisted Bernevig-Hughes-Zhang model by tuning its moiré periodic mass term. These results give a feasible method to evaluate a topological invariant for all occupied bands of a moiré material, and contribute to the design of topological moiré materials and devices.

\end{abstract}

\pacs{Valid PACS appear here}
\maketitle



\section{Introduction \label{sec:intro}}
In recent years, the study of twisted bilayer systems, or moiré materials, has become one of the most active fields in materials science \cite{cao2018correlated,cao2018unconventional,po2018origin,yankowitz2019tuning,sharpe2019emergent,jiang2019charge,polshyn2019large,lu2019superconductors,kerelsky2019maximized,choi2019electronic,xie2019spectroscopic,cao2020strange,serlin2020intrinsic,stepanov2020untying,saito2020independent,hunt2013massive,dean2013hofstadter,koshino2018maximally,moon2012energy,brihuega2012unraveling,wang2021moire,cano2021moire,carr2020electronic,kennes2021moire,tong2017topological,fujimoto2020topological,fujimoto2022perfect,kariyado2022twisted}. As in the magic-angle twisted bilayer graphene \cite{cao2018correlated,cao2018unconventional}, moiré materials have attracted much attention as a platform for realizing various quantum phases with high tunability in parameters such as the twist angle and the filling factor. The search for topological flat bands is one direction of the study of moiré materials as a foothold to obtain a correlated topological phase, such as a fractional Chern insulator without external field \cite{repellin2020chern,ledwith2020fractional,nuckolls2020strongly,xie2021fractional,wu2021chern,san2013helical,koshino2019band}. A prime example is a fragile topology of the flat band of the magic-angle twisted bilayer graphene \cite{po2019faithful,ahn2019failure,Song2019all}. In many other materials, flat moiré bands with various non-trivial topology, such as the Chern number and valley Chern number, have been proposed \cite{lian2020flat,wu2019topological,claassen2022ultra}. In these studies, the topology of the bands is evaluated with Wilson loop spectra.
\begin{figure}
    \centering
    \includegraphics[width=0.45\textwidth]{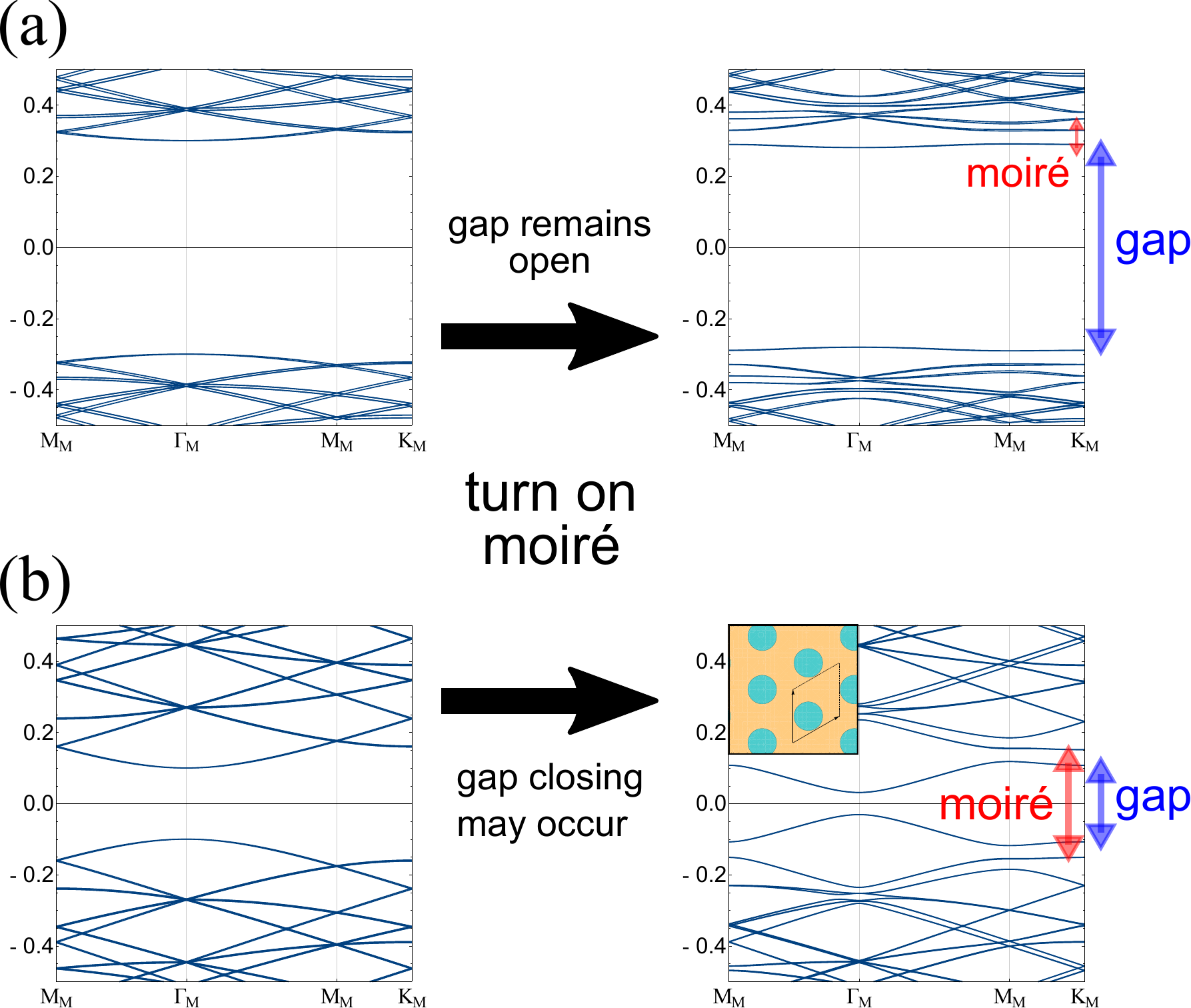}
    \caption{Schematic pictures of moiré materials with large and small moiré modulations compared with their original band gap. (a) Case where the original band gap (blue arrow in the right end) in the untwisted bilayer is larger than the amplitude of the moiré modulation (red arrow). The left is a band dispersion when the moiré modulation is neglected, and the right is when it is turned on. (b) Case where the original band gap in the untwisted bilayer is smaller than or comparable with the amplitude of the moiré modulation. In this case, a gap-closing may occur when the moiré modulation is gradually turned on.}
    \label{fig:Moiretopotrans}
\end{figure}

Generally, when a band has a nontrivial topology, a corresponding edge state appears if an edge is truncated \cite{hasan2010colloquium}. Typically, when occupied bands of an insulator have a nontrivial topological invariant, conducting edge states appear in the gap on the Fermi level. The analysis of these edge states is also important for understanding the topological properties of the system. However, the Wilson loop evaluation for a single band cannot determine which gap the edge state appears, above or below a non-trivial band. This issue is solved for normal (non-moiré) materials by evaluating all occupied bands because edge states never appear in the lower end of the occupied bands.
However, for moiré materials, it is not feasible to evaluate all occupied bands due to the huge number of occupied bands. A low-energy approximation used in moiré materials also makes the all-band evaluation impossible because it neglects bands far from the Fermi level. This obstacle in the edge state analysis has not been a problem in previous moiré materials, such as twisted bilayer MoS$_2$ family, because they have a large band gap at the Fermi level \cite{lian2020flat,wu2019topological,claassen2022ultra}\footnote{Or, it may not really matter which gap the edge states appear in because the Fermi level can be easily tuned.} that makes it easy to determine the presence or absence of edge states at the Fermi level. If the band gap at the Fermi level of the untwisted bilayer material is larger than the amplitude of moiré modulation, the presence or absence of edge states at the Fermi level of the moiré material is the same as that of the untwisted bilayer material. The reason is explained in Fig. \ref{fig:Moiretopotrans}(a). If we make a moiré supercell hypothetically neglecting the moiré modulation, the band folding by the moíre superlattice does not change a topological invariant \footnote{This statement is correct if the topological invariant is a strong invariant, such as the $\mathbb{Z}_2$ invariant in the 2D class AII. However, some topological invariants, such as the weak $\mathbb{Z}_2$ index in the 3D class AII, is not invariant against band folding.}. Even if the moiré modulation is turned on, the gap-closing does not occur and the topological invariants are preserved if it is small enough. Although moiré materials are generally capable of large carrier doping, the properties at the Fermi level can be used as a basis to discuss the properties of band gaps around.

\begin{figure}
    \centering
    \includegraphics[width=0.45\textwidth]{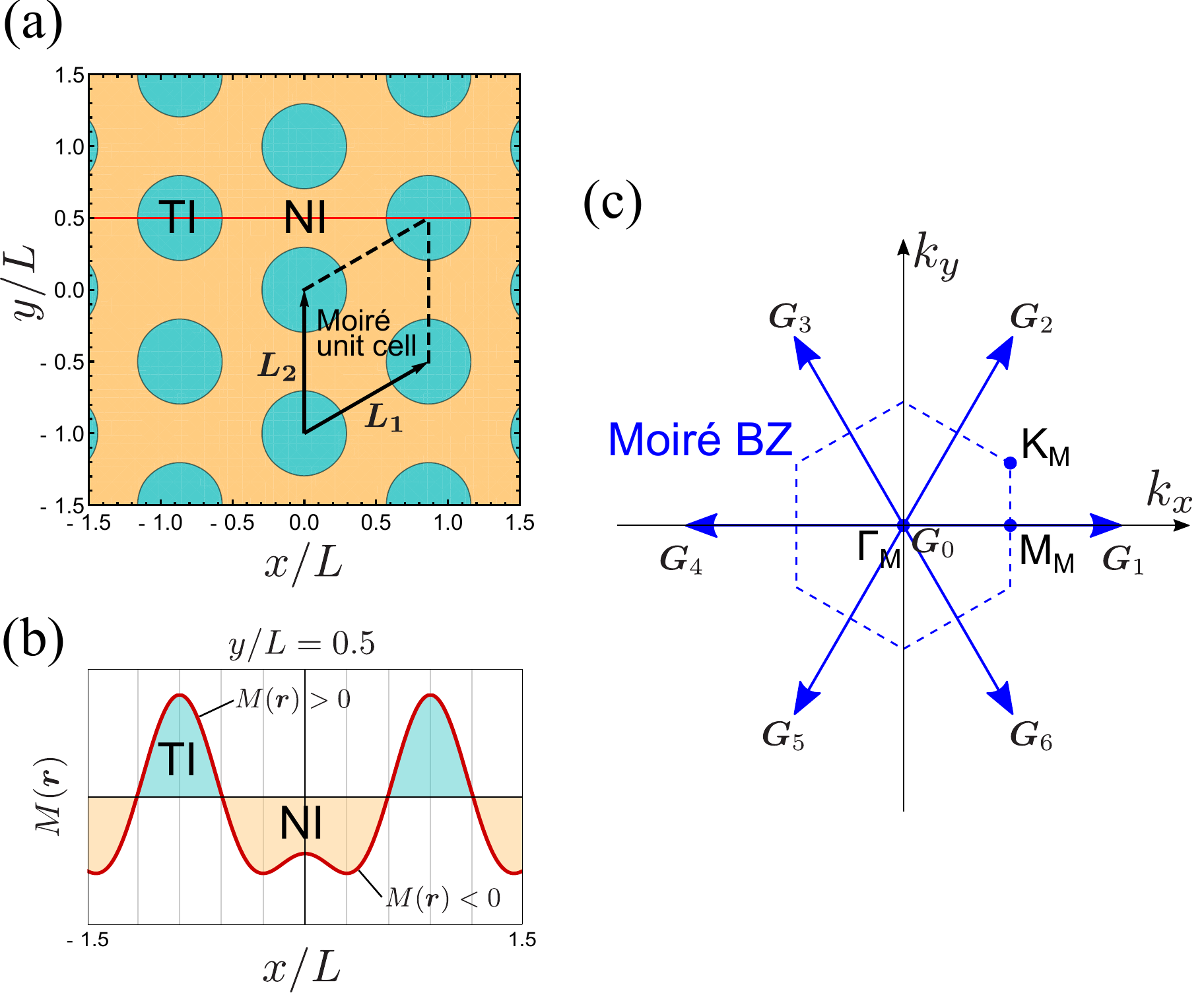}
    \caption{Schematic picture of a moiré material with TI and NI domains in the moiré unit cell. (a) Domain structure and a moiré unit cell in real space. Green regions are TI domains and an orange region is a NI domain. The domain structure is determined by the sign of $M(\bm{r]})$. (b) $M(\bm{r})$ along the line $y/L = 0.5$ (red line in (a)). (c) Reciprocal lattice vectors and a moiré BZ of the moiré system.}
    \label{fig:sche_tBHZ}
\end{figure}

However, very recently, twisted bilayer Bi$_2$(Te$_{1-x}$Se$_x$)$_3$ has been proposed \cite{tateishi2022quantum}, in which topological insulator (TI) domains and normal insulator (NI) domains coexist in the moiré unit cell (Fig. \ref{fig:sche_tBHZ}(a)). In such materials, the moiré modulation is larger than the band gap of the original untwisted bilayer (Fig. \ref{fig:Moiretopotrans}(b)). Therefore, the moiré modulation can induce a topological phase transition in the local electronic states and multiple domains of different topological phases are formed in the moiré unit cell. The previous study has shown the moiré band dispersions of these materials have a gap at the Fermi level. The question is how to determine the presence or absence of edge states in the gap at the Fermi level in such materials.
In this paper, we study a method to determine the presence or absence of edge states at the Fermi level in a moiré material with multiple topological domains in the moiré unit cell, i.e., a method to evaluate a sum of the topological invariant for all occupied bands. We show a correspondence between the topological invariant of those moiré materials at the Fermi level and the topology of the domain structure in real space by a toy model calculation. We also find a bulk-edge corresponding that is compatible with a continuous change of the truncation condition, which is specific to moiré materials.
Note that, although in a non-moiré material it is usually enough to evaluate a few valence-top bands or conduction-bottom bands to know the topological properties of the gap at the Fermi level, it does not necessarily work well for a moiré material (see Appendix \ref{sec:App_counterexample}).

This paper is organized as follows. Section \ref{sec:model} introduces a model we use to discuss the topological invariant of a moiré system with domains of two different topological phases in the moiré lattice. Section \ref{sec:method_para} presents a parameter space where we calculate a topological phase diagram and a method to determine the topological phases. Section \ref{sec:topo_phase_C3} discusses a topological phase diagram for a case when a $C_3$ symmetric mass term is assumed. Section \ref{sec:topo_phase_C2} discusses a topological phase diagram for a case when a $C_2$ symmetric mass term is assumed. Section \ref{sec:conclusion} concludes this paper. Some supplementary information is in the appendixes.

\section{model \label{sec:model}}
Here, we introduce the twisted Bernevig-Hughes-Zhang (BHZ) model, which is proposed in Ref. \cite{tateishi2022quantum}. The twisted BHZ model is used in this paper to discuss the relation between domain structure in the moiré unit cell and a topological invariant at the Fermi level. We first define the original (untwisted) BHZ model, and then we introduce a twist effect in it.

The untwisted BHZ model \cite{bernevig2006quantum} is a half-filled four-by-four (two orbitals and spin $1/2$ on each orbital) model in a two-dimensional (2D) space that describes the $\mathbb{Z}_2$ time-reversal protected topological insulator (quantum spin Hall insulator) in the class AII. We consider the $\bm{k} \cdot \bm{p}$ perturbation around the $\Gamma$ point, and the model is given as
\footnotesize
\begin{equation}
    \begin{split}
        & H_{\mathrm{BHZ}}(M,\bm{k}) \\
        &= \left( M - A k^2 \right) \tau_z \sigma_y + v k_x \tau_x \sigma_0 + v k_y \tau_y \sigma_0 \\
        &= \left(
        \begin{array}{cccc}
             & v k_- & -i(M-Ak^2) & \\
            v k_+ & & & i(M-Ak^2) \\
            i(M-Ak^2) & & & v k_- \\
             & -i(M-Ak^2) & v k_+ & \\
        \end{array}
        \right) ,
    \end{split}
    \label{eq:untwisted_BHZ}
\end{equation}
\normalsize
where $k_\pm = k_x \pm i k_y$, $k^2 = k_x^2 + k_y^2$, $A$ and $v$ are positive constant parameters. The $\sigma_{x,y,z}$, $\sigma_0$, $\tau_{x,y,z}$, $\tau_0$ are Pauli matrices and the identity matrix for the orbital and spin degree of freedom, respectively. Note that the basis is set to match the twist introduced later. This model describes a topological insulator when the mass term $M>0$ and a normal insulator when $M<0$.

Next, we define the twisted BHZ model. Generally in a moiré material with a small twist angle or small lattice constant mismatch, an atomic scale local lattice structure is approximated well with an untwisted lattice with a particular interlayer displacement \cite{bistritzer2011moire,jung2014ab}. Due to the difference in the local lattice structure, the interlayer terms have position dependence that is periodic in the moiré scale.
To introduce the twisted BHZ model, we assume that a Hamiltonian of the local structure is described by Eq.(\ref{eq:untwisted_BHZ}). The upper (lower) two rows are the spin degree of freedom on an orbital in the upper (lower) layer. It is also assumed that the mass term has a Moiré scale modulation as $M(\bm{r})$, and all the other parameters are constant. Depending on the sign of the local mass term $M(\bm{r})$, a TI (NI) domain in the moiré unit cell is defined as where $M(\bm{r})>0$ ($M(\bm{r})<0$) (Figs. \ref{fig:sche_tBHZ}(a)(b)). Because $M(\bm{r})$ is moiré-scale periodic, $H_{\mathrm{BHZ}}(M(\bm{r}),\bm{k})$ can be decomposed into components of moiré reciprocal lattice vectors $\bm{G}_l$ by Fourier transform,
\begin{equation}
    \begin{split}
        H_{\mathrm{BHZ}}(M(\bm{r}),\bm{k}) = \sum_{\bm{G}_l} e^{i \bm{G}_l \cdot \bm{r}} t_{\bm{k},\bm{k}-\bm{G}_l} .
    \end{split}
    \label{eq:BHZ_FT}
\end{equation}
By using the Fourier components $t_{\bm{k},\bm{k}-\bm{G}_l}$, the Hamiltonian of the twisted BHZ model is given as
\begin{equation}
    \begin{split}
        H = \int d \bm{k} \sum_{\alpha,\beta} \sum_{\bm{G}_l} t_{\bm{k},\bm{k}-\bm{G}_l}^{\alpha,\beta} c_{\alpha,\bm{k}}^\dagger c_{\beta,\bm{k}-\bm{G}}^{\phantom{\dagger}} ,
    \end{split}
    \label{eq:H_twistedBHZ}
\end{equation}
where $\alpha$ and $\beta$ are orbital-spin index ($\alpha, \beta = 1 \sim 4 $), $c^\dagger$ ($c$) is a creation (annihilation) operator, and $t_{\bm{k},\bm{k}-\bm{G}_l}^{\alpha,\beta}$ is a matrix element of $t_{\bm{k},\bm{k}-\bm{G}_l}$.

The untwisted BHZ model Eq.(\ref{eq:untwisted_BHZ}) has the continuous rotation symmetry but the twisted BHZ model has a lower symmetry determined by the distribution of $M(\bm{r})$. In the following, we consider two cases with different symmetries because symmetry restricts an allowed domain structure. The first one is the case where $M(\bm{r})$ has $C_{3z}$ rotation symmetry and the other is where $M(\bm{r})$ has $C_{2z}$ rotation symmetry. To discuss both in the same lattice, we assume a hexagonal unit cell for the untwisted model. Basic lattice vectors of the untwisted model are set to $\bm{a}_1 = (1,0)$ and $\bm{a}_2 = (-1/2, \sqrt{3}/2)$ although they do not explicitly appear in $\bm{k} \cdot \bm{p}$ model Eq.(\ref{eq:untwisted_BHZ}). Moiré lattice vectors are then $\bm{L}_1 = L(\frac{\sqrt{3}}{2},\frac{1}{2})$ and $\bm{L}_2 = L(0,1)$, where $L=1/ \left( 2 \sin \frac{\theta}{2} \right) $ is a moiré lattice constant for a twist angle $\theta$. In the $M(\bm{r})$ setting, a finite number of sampling points are taken and the intermediate region is interpolated by a discrete Fourier transform. Because finite sampling points are taken, finite $\bm{G}_l$ are taken into account in Eqs.(\ref{eq:BHZ_FT})(\ref{eq:H_twistedBHZ}) and higher oscillating terms are neglected. In both cases of $C_3$ symmetric $M(\bm{r})$ and $C_2$ symmetric $M(\bm{r})$, seven moiré reciprocal lattice vectors $\bm{G}_l$ ($l= 0 \sim 6$) are taken (Fig. \ref{fig:sche_tBHZ}(c)). The seven $\bm{G}_l$ are defined as $\bm{G}_0=(0,0)$ and $\bm{G}_l = \frac{4 \pi}{\sqrt{3} L} \left( \cos \frac{(l-1)\pi}{3},\sin \frac{(l-1)\pi}{3} \right)$ ($l =1 \sim 6$). The moiré Brillouin Zone (BZ) is shown in Fig. \ref{fig:sche_tBHZ}(c).

\section{calculation method and parameter space \label{sec:method_para}}

We calculate topological phase diagrams of the twisted BHZ model by tuning the position-dependent mass term $M(\bm{r})$. As described above, $M(\bm{r})$ is determined from finite sampling points, a phase diagram is given in a parameter space made by sampled mass values $\{ M_n \}$. Because the interpolation method for the intermediate area has been determined, the setting of $\{ M_n \}$ is equivalent to the setting of the domain structure. Therefore, in the parameter space $\{ M_n \}$, we can discuss the relation between the topological phases and the domain structure.

Although it is not feasible to evaluate a topological invariant for all occupied bands in a moiré system, it is possible to find points where gap closing occurs at the Fermi level even with an approximated model. In other words, a topological phase boundary can be determined. Then we determine the topological invariant for each phase. To determine the topological invariants of each phase, it is sufficient to calculate them at one point in each phase. In the parameter space, there are some points where the topological invariant is determined without numerical evaluations. For example, when $M(\bm{r})$ is constant, the topological invariant of the twisted BHZ model is the same as that of the untwisted BHZ model with the same mass value. Referring to those points, the topological invariants of each phase are explicitly determined. Note that although a gap-closing is not necessarily accompanied by a topological phase transition, in the following cases we have only one candidate of the phase boundary (gapless line) and thus there is no ambiguity.

\section{Topological phase diagram in $C_3$ symmetric case \label{sec:topo_phase_C3}}

\subsection{Bulk topological invariant}

\begin{figure}
    \centering
    \includegraphics[width=0.45\textwidth]{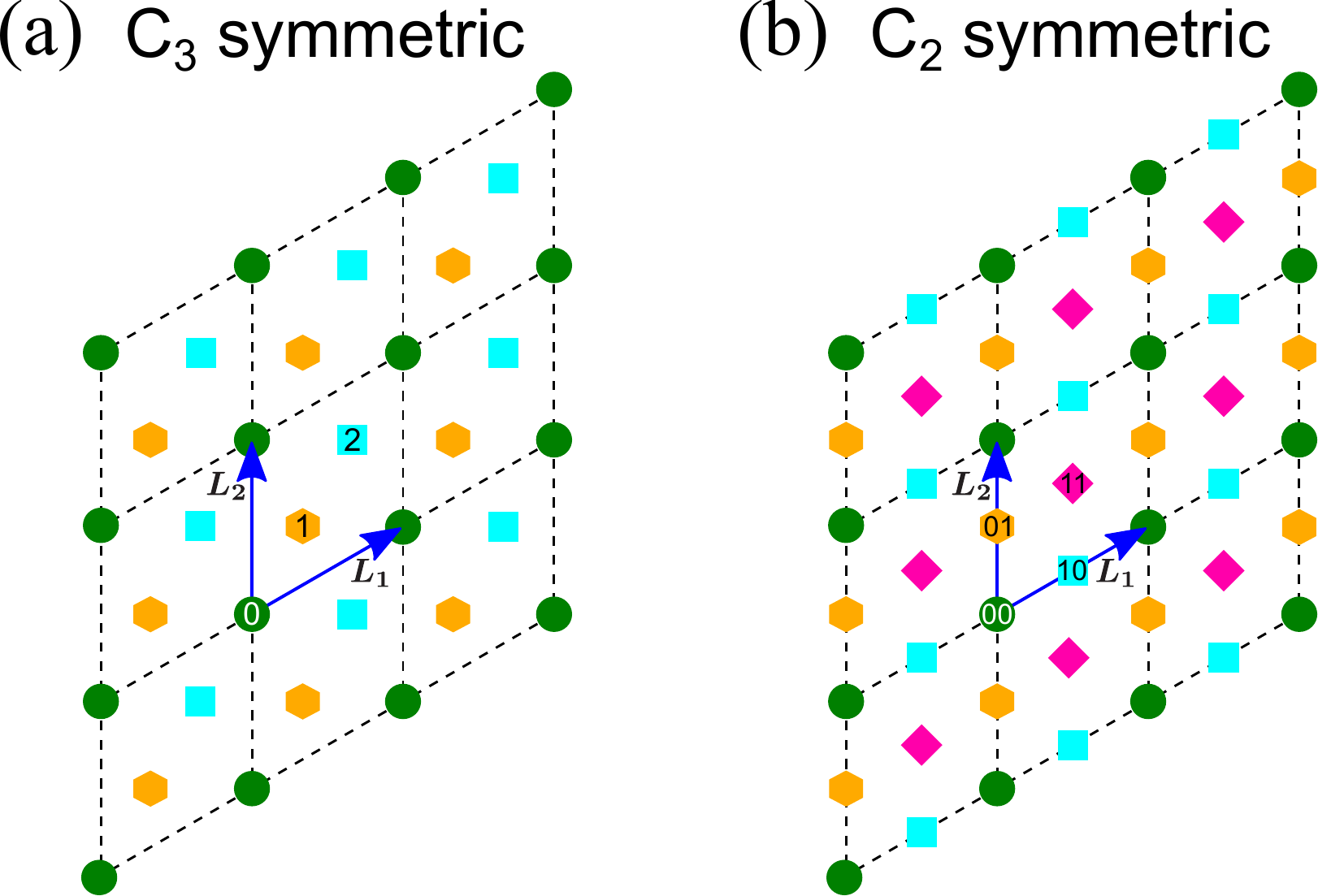}
    \caption{Sampling points to determine $M(\bm{r})$. (a) Case for a $C_3$ symmetric $M(\bm{r})$. A marker with an integer $n$ indicates a point where $M_n$ is sampled. (b) Case for a $C_2$ symmetric $M(\bm{r})$. A marker with an integer $mn$ indicates a point where $M_{mn}$ is sampled.}
    \label{fig:sampling}
\end{figure}

\begin{figure}
    \centering
    \includegraphics[width=0.45\textwidth]{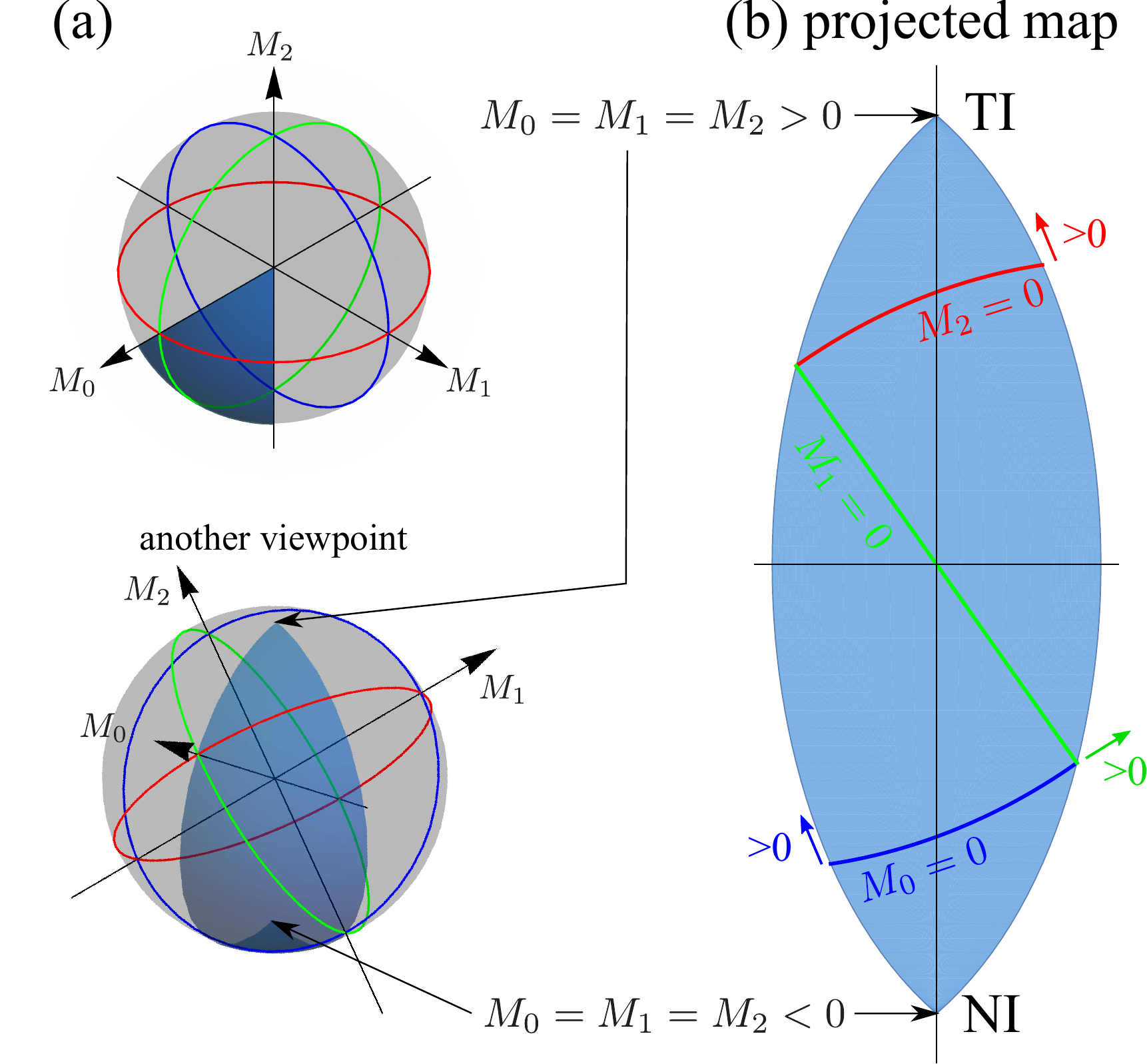}
    \caption{Parameter space ${M_0,M_1,M_2}$ for the $C_3$ symmetric case. (a) Parameter space drawn in the 3D space. Two figures are the same one viewed from (111) direction and another viewpoint. The gray sphere is where $\bar{M}=0.5$ and the blue sector is where a phase diagram is calculated. The blue, green, and red lines are $M_0=0$, $M_1=0$, and $M_2=0$, respectively. (b) Projection of the blue sector in (a) by the Lambert azimuthal equal-area projection.}
    \label{fig:C3paraspace}
\end{figure}

To set a $C_3$ symmetric $M(\bm{r})$, we take three sampling points $\bm{r}_n = \frac{n}{3} (\bm{L}_1 + \bm{L}_2)$ ($n=0,1,2$) in the moiré unit cell (Fig. \ref{fig:sampling}(a)) \footnote{We can take more sampling points to set a more detailed distribution of $M(\bm{r})$. When more sampling points are taken, we need to take in more $\bm{G}_l$ in the calculation. Here, we assume the most simplest case with three sampling points.}. In this case, the obtained moiré lattice belongs to the layer group No. 67 ($p312$). The mass values at the sampling points are defined as $M_n$. To obtain a $C_3$ symmetric twisted BHZ model, we assume equivalences between the seven Fourier components $t_{\bm{k},\bm{k}-\bm{G}_l}$ ($l=0 \sim 6$) as
\begin{equation}
    \begin{split}
        t_{\bm{k},\bm{k}-\bm{G}_1} &= t_{\bm{k},\bm{k}-\bm{G}_3} = t_{\bm{k},\bm{k}-\bm{G}_5}, \\
        t_{\bm{k},\bm{k}-\bm{G}_2} &= t_{\bm{k},\bm{k}-\bm{G}_4} = t_{\bm{k},\bm{k}-\bm{G}_6}. \\
    \end{split}
\end{equation}
With these relations, the seven $t_{\bm{k},\bm{k}-\bm{G}_l}$ are uniquely determined from the three sampling points. $t_{\bm{k},\bm{k}-\bm{G}_l}$ are written as
\begin{equation}
    \begin{split}
        t_{\bm{k},\bm{k}-\bm{G}_0} &= \frac{1}{3} \sum_{n=0}^2 H_{\mathrm{BHZ}}(M_n,\bm{k}), \\
        t_{\bm{k},\bm{k}-\bm{G}_l} &= \frac{1}{9} \sum_{n=0}^2 H_{\mathrm{BHZ}}(M_n,\bm{k}-\frac{\bm{G}_l}{2}) ~ e^{-i \frac{2\pi}{3} n} ~~~ (l=1,3,5), \\
        t_{\bm{k},\bm{k}-\bm{G}_l} &= \frac{1}{9} \sum_{n=0}^2 H_{\mathrm{BHZ}}(M_n,\bm{k}-\frac{\bm{G}_l}{2}) ~ e^{i \frac{2\pi}{3} n} ~~~ (l=2,4,6). \\
    \end{split}
    \label{eq:C3_t_def}
\end{equation}
Note that a correction $-\frac{\bm{G}_l}{2}$ is added in the second and third equations to recover an exact $C_3$ symmetry in the twisted model \cite{tateishi2022quantum}, although it is negligible in a small angle limit.

\begin{figure*}
    \centering
    \includegraphics[width=\textwidth]{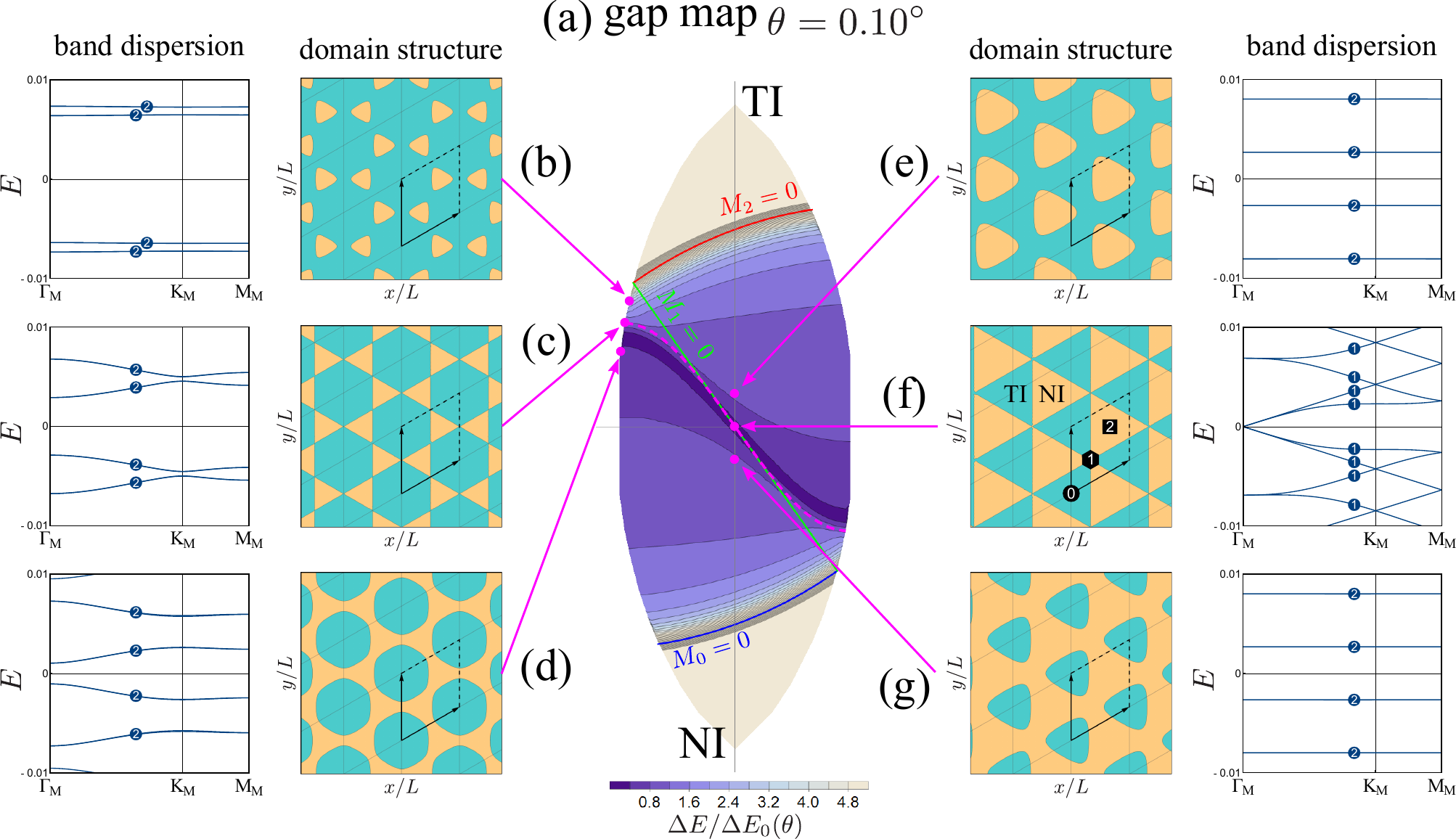}
    \caption{Calculated band gap and details in some representative points for the $C_3$ symmetric twisted BHZ model. (a) Calculated band gap for $\theta=0.10^\circ$ shown as a contour plot (gap map). The parameter space is defined in Fig. \ref{fig:C3paraspace}. The darker colored represents the smaller band gap at the Fermi level. The magenta dashed line is where the domain reconnection occurs. (b)-(g) Domain structure and band dispersion of six representative points (magenta dots in (a)). Integers in band dispersion plots indicate the degeneracy of the bands. In the counting of the degeneracy in (e) and (g), a small gap given by the interaction between domain boundaries is neglected.}
    \label{fig:C3phased1}
\end{figure*}

In a parameter space of the three sampled mass values $(M_0,M_1,M_2)$, we calculate a topological phase diagram. Before the calculation, we reduce the parameter space that we need to see by considering a symmetry between $M_n$. A cyclic exchange of the mass values $(M_0,M_1,M_2)$ $\to$ $(M_2,M_0,M_1)$ is a translation $(\bm{L}_1 + \bm{L}_2)/3$ of the moiré lattice. The time reversal protected $\mathbb{Z}_2$ invariant is independent of this translation and thus they give the same invariant. Further, exchanging two of them, for example $(M_0,M_1,M_2)$ $\to$ $(M_0,M_2,M_1)$, works as the inversion operation. Although the moiré lattice does not have the inversion symmetry, the inversion image has the same topological invariant as the original one. From the above two symmetries in $(M_0,M_1,M_2)$, the parameter space is reduced to 1/6. The origin $(0,0,0)$ gives a constant mass $M(\bm{r})=0$ that is not of our interest, so we restrict the parameters on a sphere in the parameter space to avoid the origin as
\begin{equation}
    M_{0}^2 + M_{1}^2 + M_{2}^2 =  \bar{M}^2~ =\mathrm{constant}.
    \label{eq:C3mass_sphere}
\end{equation}
Setting a north pole at where $M_0=M_1=M_2>0$ and a south pole at $M_0=M_1=M_2<0$, we calculate a phase diagram on a 1/6 sector in the longitude direction (Fig. \ref{fig:C3paraspace}). In the region above the red $M_2=0$ line (below the blue $M_0=0$ line), all $M_n$ are positive (negative) and thus the whole moiré unit cell belongs to the TI (NI) domain, while TI and NI domains coexist in the moiré unit cell in the middle region between the two lines $M_2=0$ and $M_0=0$.

As described at the beginning of section \ref{sec:method_para}, we calculate a band gap at the Fermi level and find a gap closing point in the parameter space to determine a topological phase boundary. Generally in a noncentrosymmetric 2D system, the gap closing occurs in a generic momentum in BZ \cite{murakami2007phase,murakami2007tuning}. However, when additional crystalline symmetry is present, it can be restricted to high symmetry lines \cite{yu2020piezoelectricity,tateishi2022quantum}. In the case of $C_3$ symmetric twisted BHZ model (layer group No.67), there are in-plane $C'_2$ rotation axes and the gap closing always occurs on $C'_2 {\cal T}$ invariant lines, which is $\Gamma_M$-K$_M$-M$_M$ lines. Therefore, to find a phase transition point we calculate the minimum direct gap on the $C'_2 {\cal T}$ invariant line
\begin{equation}
    \Delta E = \mathrm{min} \left\{ E_{\mathrm{LCB}}(\bm{k}) - E_{\mathrm{HVB}}(\bm{k}) \left| \bm{k} \in (\Gamma_M-\mathrm{K}_M-\mathrm{M}_M) \right. \right\}
\end{equation}
in the parameter space, where LCB stands for the lowest conduction band and HVB stands for the highest valence band.

Setting the parameters in Eqs.(\ref{eq:untwisted_BHZ}) and (\ref{eq:C3mass_sphere}) as $v=1$, $A=1$, and $\bar{M}=0.5$, we calculated $\Delta E$ in the parameter space defined in Fig. \ref{fig:C3paraspace}. In Fig. \ref{fig:C3phased1}, an obtained gap map for twist angle $\theta=0.10^\circ$ (Fig.\ref{fig:C3phased1}(a)) and domain structures of some representative points (Figs. \ref{fig:C3phased1}(b)-(g)) are shown. Fig. \ref{fig:C3phased1}(a) is a contour plot of $\Delta E$. Note that the gap is normalized with a twist angle-dependent factor
\begin{equation}
   \Delta E_0 (\theta) = \frac{v \left| \bm{G}_1 \right|}{2} = v \frac{4 \pi}{\sqrt{3}} \sin \frac{\theta}{2},
   \label{eq:gap_normal}
\end{equation}
which is a typical gap value for a case where TI (or NI) domains are isolated in the moiré unit cell (see Appendix \ref{sec:App_gap_normal} for its derivation) \cite{tateishi2022quantum}. The darker color represents where the gap is smaller. It can be seen that a gapless line goes from the lower right to the upper left. The upper (bottom) end of the parameter space is a point with $M_0=M_1=M_2>0$ ($<0$), and at that point, the moiré modulation terms vanish and thus the moiré bands are given by just folding a band of the untwisted system. Because the band folding does not change the $\mathbb{Z}_2$ topological invariant, the upper (bottom) end belongs to a TI (NI) domain. Because there is only one gapless line in the gap map, we can determine that it is the phase boundary and the upper (lower) phase is a TI (NI) phase. Next, we discuss what determines the shape of the domain boundary by reference to the domain structure in the moiré unit cell. For six representative points (magenta dots in Fig. \ref{fig:C3phased1}(a)), the domain structure in the moiré unit cell and the band dispersion are shown in Figs. \ref{fig:C3phased1}(b)-(g). The numbers in the band dispersion figures are numbers of degenerate bands including the spin degree of freedom. First, we focus on (e), (f), and (g). In (e), the TI domain is connected over the moiré lattice but the NI domains are disconnected from each other. This case belongs to the TI phase as a moiré system. Conversely, in (g), the TI domains are disconnected but the NI domains are connected. This case belongs to the NI phase as a moiré system. In the case of (f), where $M_1=0$, both domains are triangles and are touching at where $M_1$ is sampled, and the moiré system becomes gapless as shown in the right of (f). Because the inversion symmetry is broken in these cases, Rashba splitting appears clearly in the band dispersion in (f). However, in (e) and (g), the splitting is unnoticeable due to a negligible interaction between helical edge states on the domain boundary of isolated domains. Next, we focus on (b), (c), and (d), which are calculated on the left edge of the parameter space. The left edge is a line where $M_1=M_2$ is satisfied, and thus the moiré system has the $C_6$ symmetry and the (approximate) inversion symmetry as the domain structure indicates (detailed discussion about the symmetry is given in Appendix \ref{sec:App_sym_of_moire}). Due to the inversion symmetry, there is no Rashba splitting in the band dispersion in (b), (c), and (d). Also in these cases, when the TI (NI) domain is connected as (b) (as (d)), the moiré system is TI (NI). The calculated domain structures suggest a correspondence between the domain connectivity and the topological invariant of the moiré system. In Fig. \ref{fig:C3phased1}(a), a line where domain reconnection occurs (magenta dashed line in Fig. \ref{fig:C3phased1}) is given and it is qualitatively coincident with the gapless line (the dark colored region in (a)). It is noteworthy that the connectivity of the domains determines the topological phase of the system, not the size of the area of each domain. The case of (d) is a good example, where the system is NI although roughly 62\% of the moiré unit cell is the TI domain.

\begin{figure}
    \centering
    \includegraphics[width=0.45\textwidth]{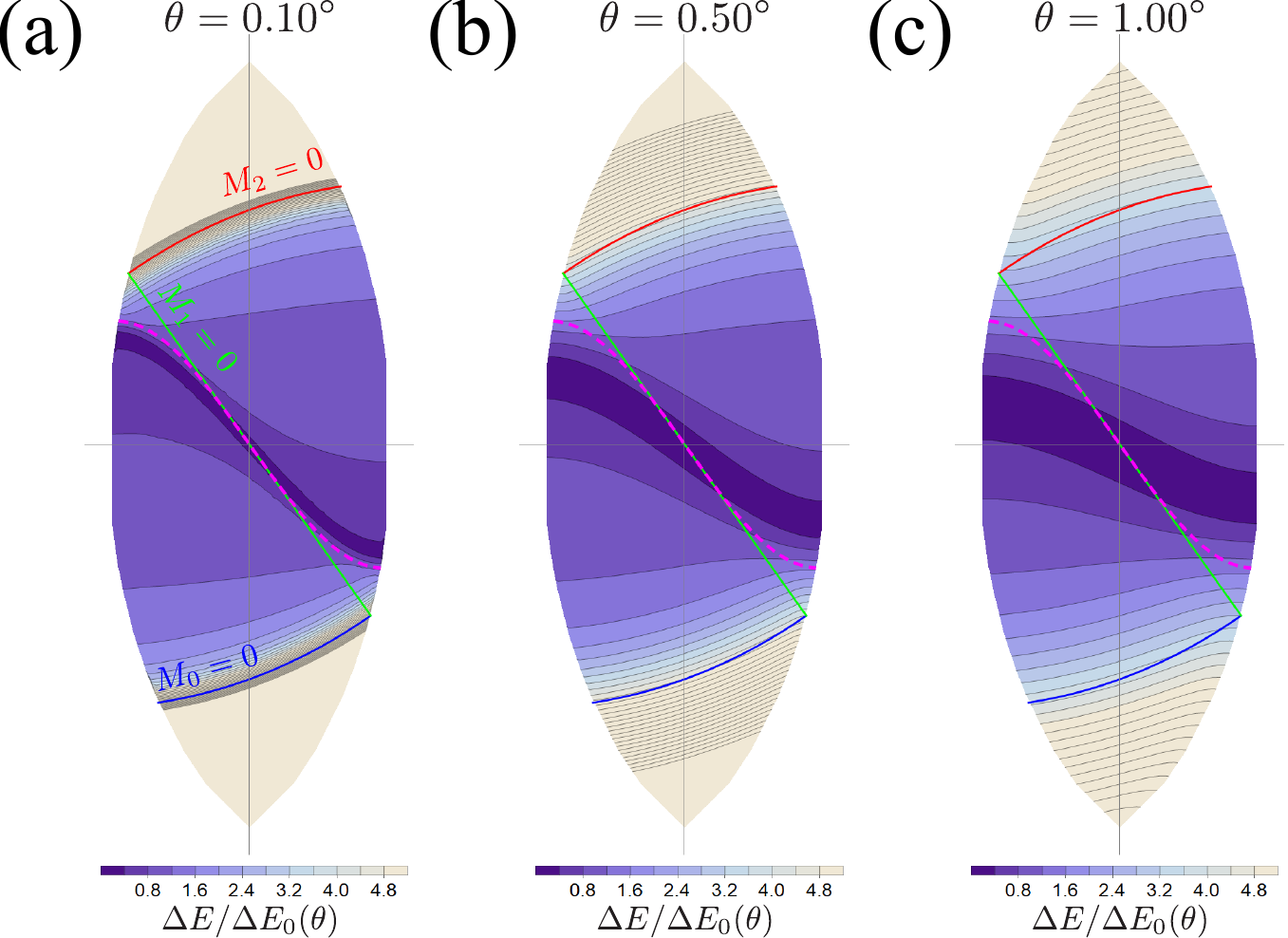}
    \caption{Twist angle dependence of the gap map for the $C_3$ symmetric twisted BHZ model. (a) $\theta=0.10^\circ$ (b) $\theta=0.50^\circ$, and (c) $\theta=1.00^\circ$, respectively.}
    \label{fig:C3angle_depend}
\end{figure}

As shown in Fig. \ref{fig:C3phased1}, the shape of the phase boundary is understood mainly by the domain connectivity. However, on the side edge, the gapless point is slightly shifted from the point where domain reconnection occurs. On the left edge, domain reconnection occurs at a point of Fig. \ref{fig:C3phased1}(c), where $M_0 = -8 M_1$, $M_1=M_2$ is satisfied (intersection of the edge and the dashed magenta line) and Kagomé domain structure is realized (the middle left case). However, the band dispersion of Fig. \ref{fig:C3phased1}(c) is gapped and the system is TI as a moiré system. This difference is explained by an effect of the interaction between the helical edge states on the domain boundary. Fig. \ref{fig:C3angle_depend} shows a twist angle dependence of the gap map. As the twist angle increases from Fig. \ref{fig:C3angle_depend}(a) to Fig. \ref{fig:C3angle_depend}(c), the difference between the gapless line and the domain reconnection line becomes more significant. Because the moiré unit cell becomes smaller for a larger twist angle case, the width of the helical edge state on the domain boundary becomes wider compared to the moiré lattice scale, as reported in Ref. \cite{tateishi2022quantum}. As a result, interactions are no longer negligible in cases where domain boundaries get close to each other. This interaction shifts the parameter at which the reconnection of the helical edge states occurs from that at which the domain reconnection occurs. This is a source of the shift of the gapless line from the domain reconnection line. This effect is not found in the case of Fig. \ref{fig:C3phased1}(f) because the TI and NI domains are symmetric and the edge states cannot choose which domain to go through.

\subsection{Edge state}
\begin{figure}
    \centering
    \includegraphics[width=0.45\textwidth]{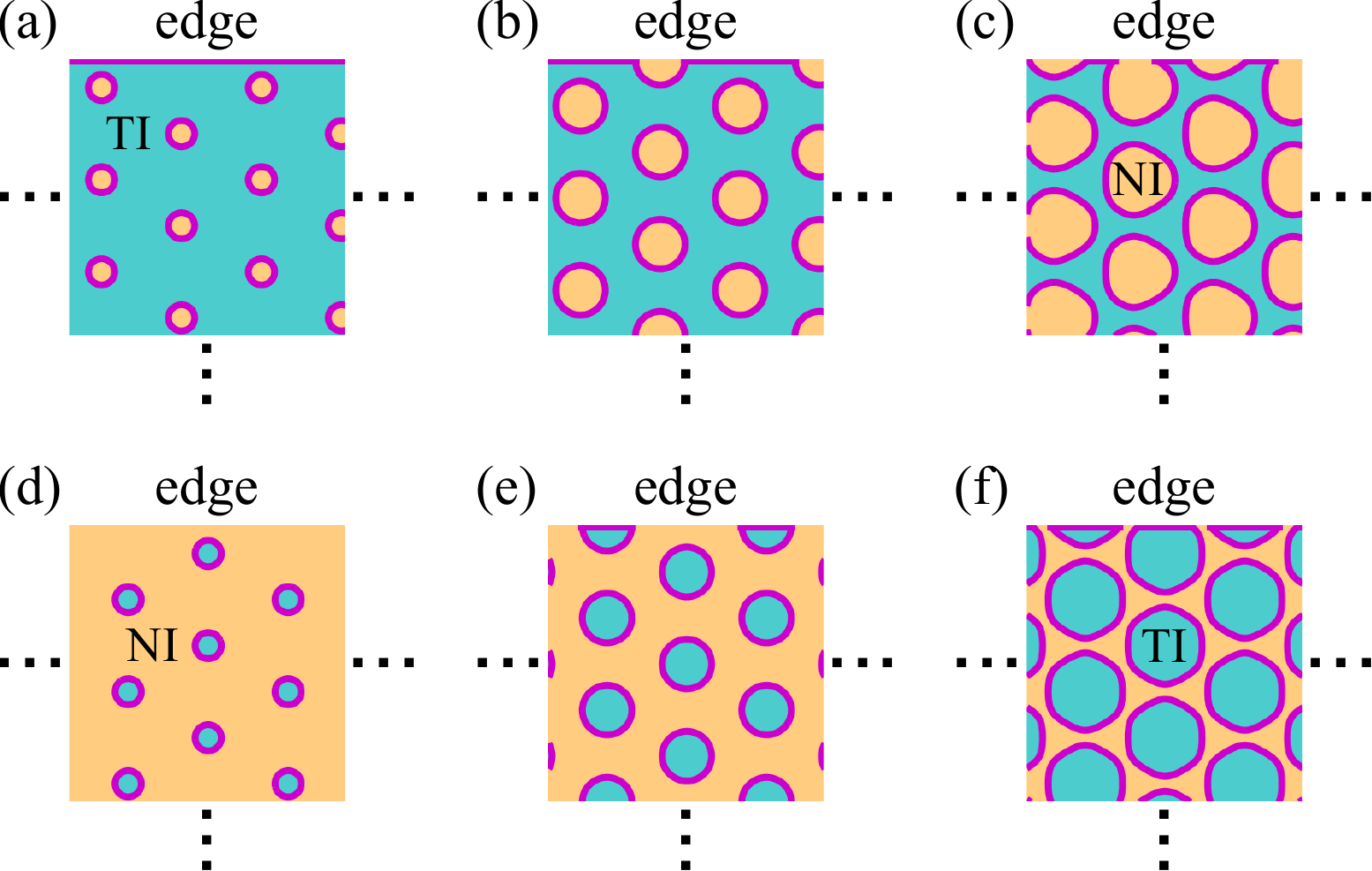}
    \caption{Domain connectivity and edge states. (a)(b)(c) When the TI domain is connected, there is a connected edge state along the edge from one end to the other end. Even if some parts of the edge are in the NI domains, the edge state is connected though the domain boundary as shown in (b) and (c). (d)(e)(f) When the NI domain is connected, there cannot be a connected edge state along the edge.}
    \label{fig:C3edge}
\end{figure}

The obtained correspondence between the topological invariant and domain connectivity is consistent with the existence of edge states. In Fig. \ref{fig:C3edge}, examples of some domain structures and truncation conditions are shown. When an edge cuts a TI domain, the edge is locally considered a boundary between a TI and the vacuum and thus there must be an edge state (Fig. \ref{fig:C3edge}(a)). Combined with edge states on the TI and NI domain boundary, we can see there always is a connected edge state along the edge when the system has a TI-connected domain structure (Figs. \ref{fig:C3edge}(a)-(c)). Although NI domains can be cut in some parts of the edge, a connected edge state exists through the TI-NI domain boundary (Figs. \ref{fig:C3edge}(b)-(c)). In contrast, there cannot be a connected edge state along the edge when the system has a NI-connected domain structure, i. e., a TI-disconnected domain structure (Figs. \ref{fig:C3edge}(d)-(f)). Even if most parts of the edge cut TI domains, the edge state is disconnected by the connected NI domain (Fig. \ref{fig:C3edge}(f)). These results show a clear correspondence between the bulk topological invariant and the edge state, and the connected edge state is a corresponding edge state of the moiré topological insulator. It is noteworthy that the existence of the connected edge state is independent of a truncation condition. This is consistent with the fact that the $\mathbb{Z}_2$ bulk invariant in the class AII generally guarantees the existence of helical edge states regardless of the truncation condition. In a normal (non-moiré) crystal, only discretized truncation conditions are allowed because one cannot divide an atom, while in a moiré system, effectively continuous truncation conditions are allowed especially in the small twist angle limit. The obtained bulk-edge correspondence is consistent with this moiré-specific continuous truncation conditions.

\section{Topological phase diagram in $C_2$ symmetric case \label{sec:topo_phase_C2}}
In Sec.\ref{sec:topo_phase_C3}, we found a correspondence between the topological invariant and the domain structure. In the $C_3$ symmetric case, one of the TI and NI domains is connected and the other is disconnected. This setup suits the $\mathbb{Z}_2$ nature of the topological phase of the class AII. However, in a general case, the domain structure is not necessarily classified into two classes. For example, a stripe pattern of TI and NI domains is allowed, in which both domains are connected in one direction but disconnected to the other perpendicular direction. However, we do not know any anisotropic topological phase in 2D like the weak TI in three-dimensional (3D) space. In the following, we discuss what topological phase a system with such stripe domain structures belongs to. To reduce the number of sampling points while allowing a stripe pattern, we assume $C_2$ symmetry. 

\subsection{Bulk topological invariant}

\begin{figure*}
    \centering
    \includegraphics[width=\textwidth]{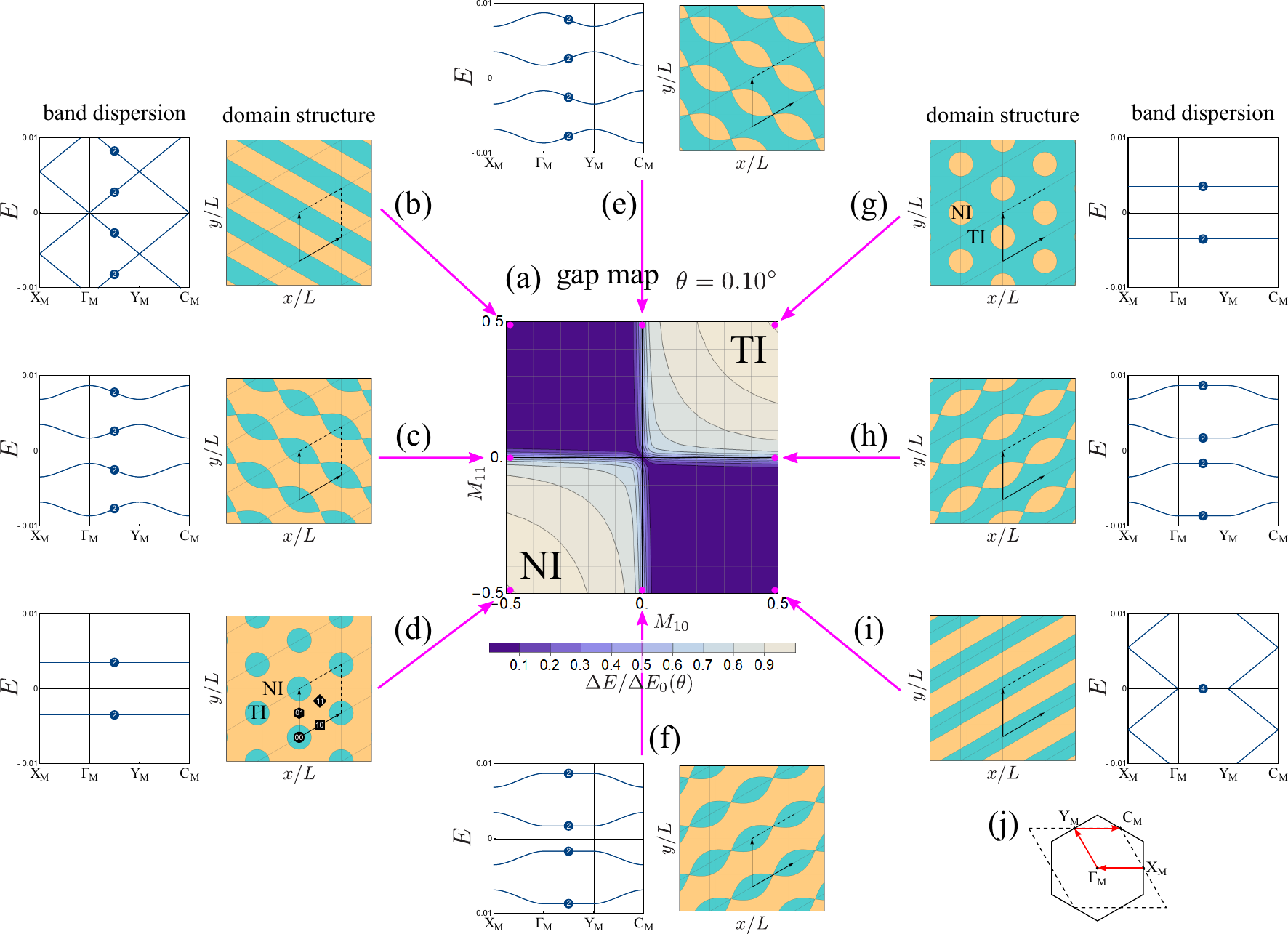}
    \caption{Calculated band gap and details in some representative points for the $C_2$ symmetric twisted BHZ model. (a) Calculated band gap for $\theta=0.10^\circ$ shown as a contour plot. The parameter space is ${M_{10},M_{11}}$ and the other two are fixed as $M_{00}=0.5$ and $M_{01}=-0.5$. The dark-colored region is where the band gap at the Fermi level closes. (b)-(i) Domain structure and band dispersion of eight representative points (magenta dots in (a)). Integers in band dispersion plots indicate the degeneracy of the bands. (j) Path of momentum space used to plot the band dispersions in (b)-(i).}
    \label{fig:C2phase}
\end{figure*}

To set a $C_2$ symmetric $M(\bm{r})$, we take four sampling points $\bm{r}_{mn} = \frac{1}{2} \left( m \bm{L}_1 + n \bm{L}_2 \right)$ ($m,n=0,1$) in the moiré unit cell (Fig. \ref{fig:sampling}(b)). In this case, the obtained moiré lattice belongs to the layer group No. 3 ($p112$). However, in our model, the inversion symmetry is approximately recovered due to the small angle approximation and too simple untwisted models in the sampling points (see Appendix \ref{sec:App_sym_of_moire} for details). The mass values at the sampling points are defined as $M_{mn}$. To obtain a $C_2$ symmetric twisted BHZ model, we assume equivalence between the seven Fourier components $t_{\bm{k},\bm{k}-\bm{G}_l}$ as
\begin{equation}
    \begin{split}
        t_{\bm{k},\bm{k}-\bm{G}_1} &= t_{\bm{k},\bm{k}-\bm{G}_4}, \\
        t_{\bm{k},\bm{k}-\bm{G}_2} &= t_{\bm{k},\bm{k}-\bm{G}_5}, \\
        t_{\bm{k},\bm{k}-\bm{G}_3} &= t_{\bm{k},\bm{k}-\bm{G}_6}. \\
    \end{split}
\end{equation}
With these relations, the seven $t_{\bm{k},\bm{k}-\bm{G}_l}$ are uniquely determined from the four sampling points. $t_{\bm{k},\bm{k}-\bm{G}_l}$ are written as
\begin{equation}
    \begin{split}
        t_{\bm{k},\bm{k}-\bm{G}_0} &= \frac{1}{4} \sum_{\substack{m=0,1\\n=0,1}} H_{\mathrm{BHZ}}(M_{mn},\bm{k}), \\
        t_{\bm{k},\bm{k}-\bm{G}_l} &= \frac{1}{8} \sum_{\substack{m=0,1\\n=0,1}} H_{\mathrm{BHZ}}(M_{mn},\bm{k}-\frac{\bm{G}_l}{2}) ~e^{i \pi n} ~~(l=1,4), \\
        t_{\bm{k},\bm{k}-\bm{G}_l} &= \frac{1}{8} \sum_{\substack{m=0,1\\n=0,1}} H_{\mathrm{BHZ}}(M_{mn},\bm{k}-\frac{\bm{G}_l}{2}) ~e^{i \pi (m+n)} ~~(l=2,5), \\
        t_{\bm{k},\bm{k}-\bm{G}_l} &= \frac{1}{8} \sum_{\substack{m=0,1\\n=0,1}} H_{\mathrm{BHZ}}(M_{mn},\bm{k}-\frac{\bm{G}_l}{2}) ~e^{i \pi m} ~~(l=3,6). \\
    \end{split}
    \label{eq:C2_t_def}
\end{equation}
A correction $-\frac{\bm{G}_l}{2}$ is added as explained in Sec.\ref{sec:topo_phase_C3}.

Next, we determine a parameter space in which to calculate a topological phase diagram. The parameter space is in principle four-dimensional space of $M_{mn}$. However, now we are interested in the topological invariant of stripe cases. Therefore, we fix two of them as $M_{00}=0.5$ and $M_{01}=-0.5$ and use the other two as axes of a phase diagram. In this parameter setting, we can see TI-connected, NI-connected, and stripe domain structures in the parameter space.

In the parameter space $(M_{10},M_{11})$, we calculate a band gap and find a gap closing point to determine a topological phase boundary. In this case, approximate inversion symmetry is recovered 
as described above. Therefore, a gap closing of a topological phase transition occurs on a time-reversal invariant momentum (TRIM). Then, a band gap $\Delta E$ is defined as
\begin{equation}
    \Delta E = \min \left\{ E_{\mathrm{LCB}}(\bm{k}) - E_{\mathrm{HVB}}(\bm{k}) \left| \bm{k} \in \mathrm{TRIM} \right. \right\} .
\end{equation}

Setting the parameters in Eq.(\ref{eq:untwisted_BHZ}) as $v=1$ and $A=1$, we calculate $\Delta E$ in the parameter space $(M_{10},M_{11}) \in [-0.5,0.5] \times [-0.5,0.5] $. An obtained gap map for twist angle $\theta = 0.10 ^\circ$ is shown in Fig. \ref{fig:C2phase}(a) as a contour plot. Note that the gap is normalized with $\Delta_0(\theta)$. In Figs. \ref{fig:C2phase}(b)-(i), domain structures and band dispersions are also shown for some representative points in the parameter space. The band dispersion figures are plotted along a path shown in (j). In the cases of (d) and (g), the domain structures are NI- and TI-connected, respectively. In the cases of (b) and (i), the domain structures are stripe structures. As shown in the gap map (a), parameter regions of stripe domain structure, \{$M_{10}>0$ and $M_{11}<0$\} or \{$M_{10}<0$ and $M_{11}>0$\}, are gapless as a moiré system (dark color). This is a reasonable result because there should be helical edge states on the parallel domain boundaries. When the twist angle is small enough and the helical edge states are ideally isolated, the helical edge states give gapless linear dispersion. In the band dispersion of (b), we can see a clear linear dispersion. The double degeneracy is due to the presence of two domain boundaries in the moiré unit cell . In the other stripe case of (i), we can see flat zero-energy bands in the $\Gamma_M$-Y$_M$ line. This is because the direction of the $k$ path of that part is perpendicular to the direction of the stripe and thus the helical edge states cannot run in that direction. On the two axes, (c), (e), (f), and (h) that are on $M_{10}=0$ or $M_{11}=0$, a domain reconnection occurs from a TI-connected (or NI-connected) to a stripe domain structure. On the domain reconnection points, the domain boundaries singularly touch each other and thus the system is gapped due to the interaction between helical edge states. From these results, we have obtained a qualitative picture of the topological phase diagram in a small twist-angle case. We know the case of (g) (the case of (d)) with a TI-connected (NI-connected) domain structure is a TI (NI) as a moiré system as shown in Sec. \ref{sec:topo_phase_C3}. The cases with stripe domain structures are gapless and thus the topological invariant is not well-defined.

\begin{figure}
    \centering
    \includegraphics[width=0.45\textwidth]{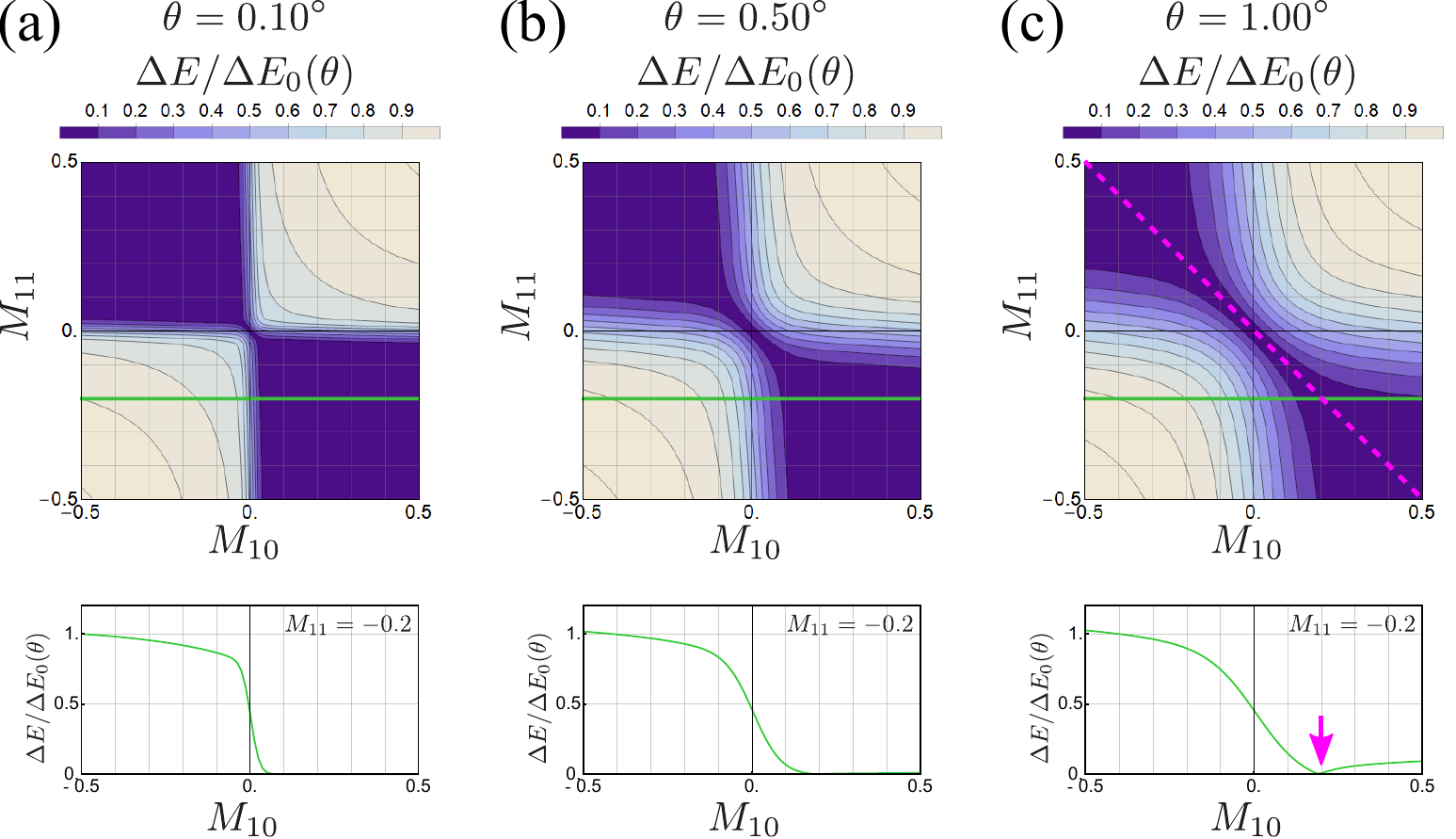}
    \caption{Twist angle dependence of the gap map for the $C_2$ symmetric twisted BHZ model. From left to right: one for $\theta=0.10^\circ$, $\theta=0.50^\circ$, and $\theta=1.00^\circ$, respectively. In the bottom of each one, the normalized gap on the $M_{11}=-0.2$ (green line) is shown. For $\theta=1.00^\circ$, a line (point) where the band gap exactly closes is shown as a megenta dashed line (magenta arrow).}
    \label{fig:C2angle_depend}
\end{figure}

However, in the general theory of the topological phase transition in the 2D class AII, a phase transition occurs at a point in a single parameter tuning, i. e., the gapless points should be a line with no width in a 2D parameter space \cite{murakami2007phase,murakami2007tuning}. Next, we explain why gapless regions can be obtained contrary to the general theory. The gapless region is a result of an additional condition that interactions between neighboring helical edge states can be neglected in the small angle limit. It is confirmed by examining the twist angle dependence of the gap map (Fig. \ref{fig:C2angle_depend}). In the bottom of Fig. \ref{fig:C2angle_depend}, The band gap on the $M_{11}=-0.2$ line is also shown. While the stripe region is (almost) gapless when $\theta=0.10^\circ$ (Fig. \ref{fig:C2angle_depend}(a)), there is a small but non-negligible gap when $\theta=1.00^\circ$ (Fig. \ref{fig:C2angle_depend}(c)). The exact gapless points in the case of $\theta=1.00^\circ$ exist on a line $M_{10}=-M_{11}$, which is a phase boundary. The small gap in the stripe region is given by an interaction between neighboring helical edge states. The location of the gapless line is determined by whether interactions across the TI or NI domain are dominant, like the Su–Schrieffer–Heeger model \cite{su1979solitons}. The interaction is always non-zero in a strict sense but, when the twist angle is small enough, each domain boundaries are distant enough in the real space and thus the interaction is negligible. In that case, the stripe region becomes a gapless region approximately even in the normalized scale by $\Delta_0(\theta)$, as shown in Fig. \ref{fig:C2angle_depend}(a). On the other hand, the gap in TI-connected (NI-connected) region is given by a finite size effect along the closed loop of the domain boundary. Therefore, the band gap in those regions is roughly 1 and independent of the twist angle in the normalized scale. The general theory of the topological phase transition assumes all perturbations allowed in the symmetry. Therefore, when we consider the correspondence to the general theory in present cases, the interactions between neighboring domain boundaries, which are allowed to exist by the symmetry of the system, must be included. When the interaction is negligible as in a small angle case, neglecting the interaction works as an additional condition and a gapless region with finite width can appear.

To summarize the result above, the property of the moiré system is still mainly determined by the domain structure in the $C_2$ symmetric case, and the case with a stripe domain structure is a gapless system. The effect of the interaction between helical edge states on the domain boundaries comes out differently, and it reduces the dimension of the gapless area to a line, as expected in the general theory of the topological phase transition. Here, whether the interaction between neighboring domain boundaries is negligible or not is judged by comparison with the typical gap value $\Delta E_0(\theta)$, which comes from the finite size effect along the closed loop of the domain boundary.

\subsection{Edge state}
\begin{figure}
    \centering
    \includegraphics[width=0.45\textwidth]{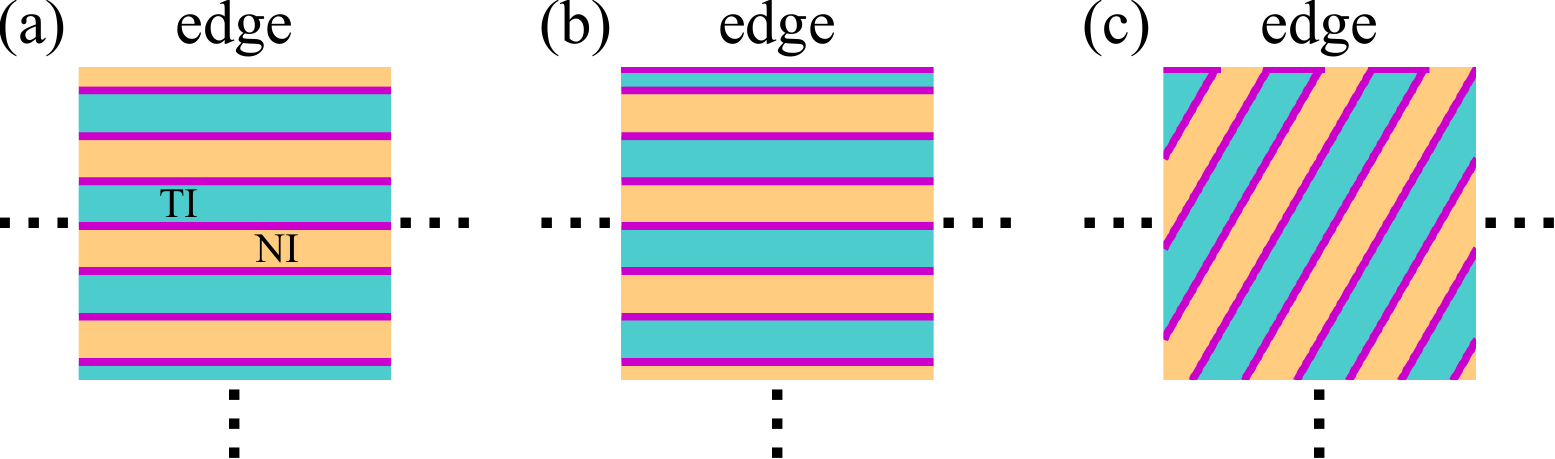}
    \caption{Stripe domain structure and edge states. (a) When the edge is parallel to the stripe and in a NI domain. (b) When the edge is parallel to the stripe and in a TI domain. In this case, there is an edge state but it is indistinguishable from the bulk transport. (c) When the edge is not parallel to the stripe.}
    \label{fig:C2edge}
\end{figure}

In the case of a stripe domain structure, the system is gapless and the edge transport is indistinguishable from the bulk transport. Figs. \ref{fig:C2edge}(a) and (b) show the cases when the edge is truncated parallel to the stripe pattern. If the edge is in the TI domain, there are helical edge states on the edge (Fig. \ref{fig:C2edge}(a)). However, there also are helical edge states on the domain boundaries in the bulk region that contribute to the gapless bulk transport. When the twist angle is small enough and the helical edge states are not interacting with each other, the helical edge states in the bulk region and edge are equivalent and no edge-specific transport is obtained. When the edge is not parallel to the stripe pattern (Fig. \ref{fig:C2edge}(c)), there is no connected edge state on the edge. In this case, the bulk region is not conducting in the direction parallel to the edge either. If the interaction between helical edge states on the domain boundary is not negligible and that across NI domains is dominant, the helical edge states get gapped by making pairs across NI domains. In this process, if the edge is in a TI domain (Fig. \ref{fig:C2edge}(b)), the helical edge states on the edge are left, and thus only the edge remains gapless. If the edge is not parallel to the stripe pattern, the disconnected parts are effectively connected by the interaction and gapless edge states are obtained.

\section{Discussion and conclusion \label{sec:conclusion}}
In this paper, we discussed the topological invariant of the twisted BHZ model at the Fermi level, in particular for the cases when TI and NI domains coexist in the moiré unit cell. As a result, we found the topological invariant of the moiré system is determined by the topology of the domain connectivity in the real space when the twist angle is small enough to neglect interactions between neighboring domain boundaries. We also found a bulk-edge correspondence that is described by the presence or absence of a connected line of edge states along the truncated edge. The obtained bulk-edge correspondence is compatible with the continuous change of the edge truncation condition that is specific to moiré materials. The effect of the interaction between neighboring domain boundaries is also discussed, and it is shown to work as a correction of the location or width of the phase boundary. Although we discussed the topological properties of the twisted BHZ model in this paper, the obtained results suggest that they can be applied to other topological phases, at least $\mathbb{Z}_2$ topological phases. These results give a method to calculate a topological invariant for cases where several topological domains coexist in the moiré unit cell due to the moiré modulation terms. The significance of the method is that it does not require a high-cost calculation such as the all-band calculation. The system we consider in this study, which has a superlattice structure of different topological domains, can be a platform for realizing novel topological systems such as a Weyl semimetal in a multilayer heterostructure proposed by Burkov and Balents \cite{burkov2011weyl}. This understanding of the topological properties of moiré materials is expected to allow us to design novel topological moiré materials with unique properties, in a way like a ``puzzle" with pieces of topological phases.

\section*{Acknowledgements}
We thank Aaron Merlin M\"uller for discussions.
This work was supported by JST CREST (Grants No. JPMJCR19T2).
M.H. was supported by PRESTO, JST (JPMJPR21Q6) and JSPS KAKENHI Grants No. 20K14390.

\appendix

\section{Example of failure in Wilson loop evaluation \label{sec:App_counterexample}}
\begin{figure}
    \centering
    \includegraphics[width=0.45\textwidth]{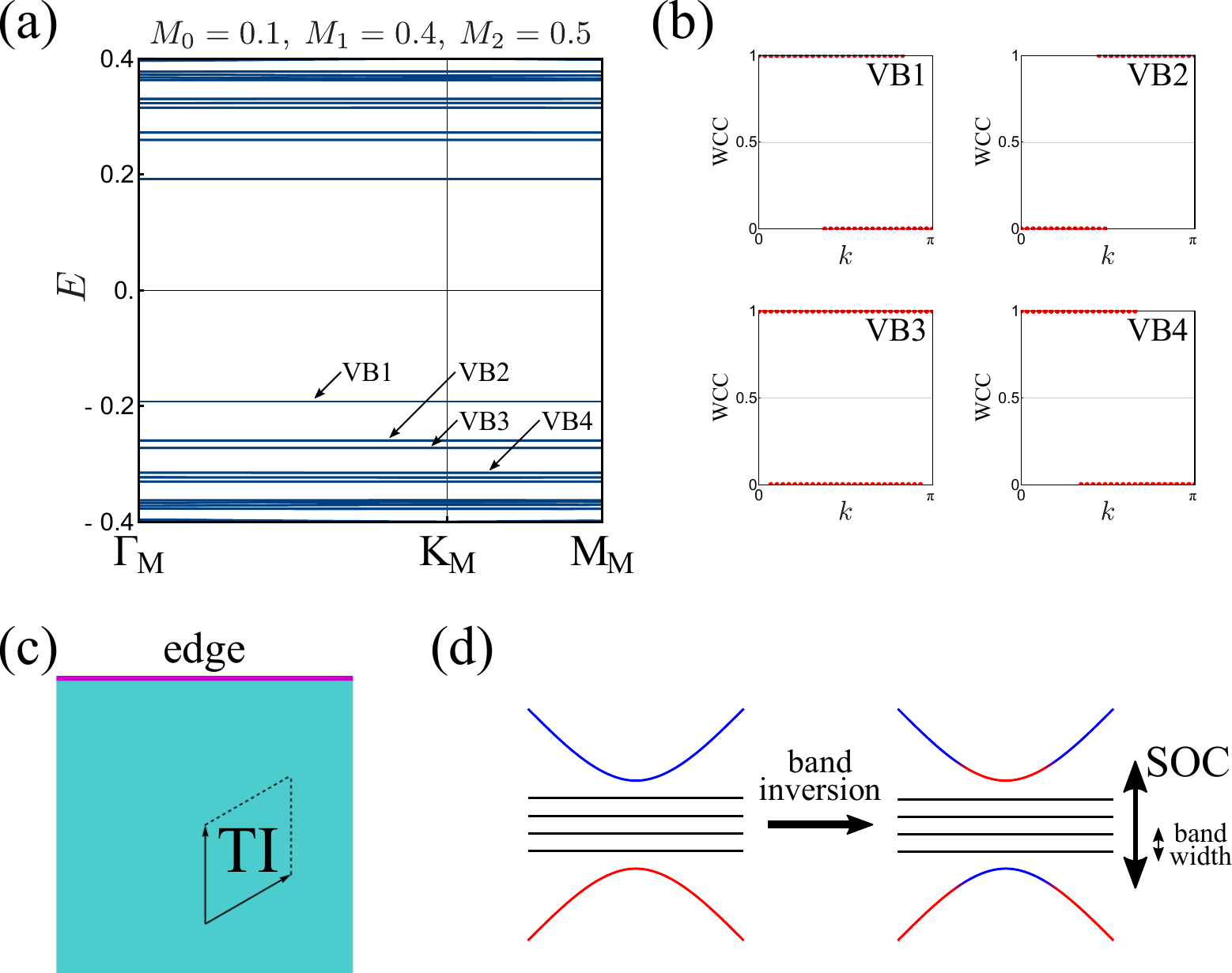}
    \caption{Example where Wilson loop evaluation of a few valence-top bands fail to capture the non-trivial property. (a) Moiré band dispersion of the $C_3$ symmetric twisted BHZ model with $M_0=0.1$, $M_1=0.4$, and $M_2=0.5$. (b) Wilson loop spectra for the four Kramers pairs of valence-top bands. (c) Domain structure and edge state. In this case, the whole moiré unit cell belongs to a TI domain and there is an edge state. (d) Schematic picture of the non-trivial band structure in this system. The band inversion occurs across several nearly flat moiré bands.}
    \label{fig:counterexample}
\end{figure}

Here, we give an example where the Wilson loop evaluation for a few valence top bands fail to capture the non-trivial topology of a moiré band structure. Fig.\ref{fig:counterexample}(a) is a band dispersion of the $C_3$ symmetric twisted BHZ model (Eqs.(\ref{eq:H_twistedBHZ} and (\ref{eq:C3_t_def})) with $M_0=0.1$, $M_1=0.4$, and $M_2=0.5$. The band dispersion is gapped and isolated nearly flat bands appear around the Fermi level. Wilson loop spectra calculated for the four Kramers pairs of nearly flat bands are shown in Fig.\ref{fig:counterexample}(b) and all of them are trivial. However, in this case, the three mass values $M_n$ are all positive and thus the whole moiré unit cell belongs to a TI domain (Fig.\ref{fig:counterexample}(c)). Therefore, there must be gapless helical edge states on the edge, even though the Wilson loop spectra are trivial for the bands around the Fermi level. This is because the non-trivial structure of this system exists in higher- or lower-energy moiré bands. When a band inversion occurs to give a TI phase, the band gap is usually determined by the energy scale of spin-orbit coupling (SOC). In a moiré material, the bandwidth generally gets quite small so that it can be far smaller than the energy scale of SOC. As a result, a band inversion can occur across some nearly flat bands (Fig.\ref{fig:counterexample}(d)), and the moiré bands around the Fermi level remain trivial. In a normal (non-moiré) material, this band inversion does not usually occur because the bandwidth is larger than the energy scale of SOC. However, it easily occurs in a moiré material, and thus evaluating a few valence top bands is not necessarily sufficient to determine the topological properties of the moiré material.

\section{Gap normalization factor \label{sec:App_gap_normal}}
The gap normalization factor $\Delta E_0 (\theta)$ in Eq.(\ref{eq:gap_normal}) is given as followings. When there are isolated TI (or NI) domains in the moiré lattice, the low-energy physics is described by edge states on the domain boundary \cite{tateishi2022quantum}. The edge state basically has linear dispersion (Dirac cone), but due to the moiré effect, the band is folded with moiré BZ and gets gapped to make nearly flat bands by the moiré modulation, which mostly comes from interlayer coupling terms. From another perspective, this flat band formation can be understood as a result of the finite size effect given by the finite length of the domain boundary that is making a closed loop. If we assume that a perfect flat band is obtained at the middle level of the original bandwidth by the moiré effect, the band gap is
\begin{equation}
   \Delta E_0 (\theta) = \frac{v \left| \bm{G}_1 \right|}{2} = v \frac{4 \pi}{\sqrt{3}} \sin \frac{\theta}{2}.
\end{equation}
Although the formation of the perfect flat band is a too simplified assumption, this value can be used as a typical energy scale of the band gap and as a normalization factor to enable a comparison between cases with different twist angles.

\begin{figure}
    \centering
    \includegraphics[width=0.45\textwidth]{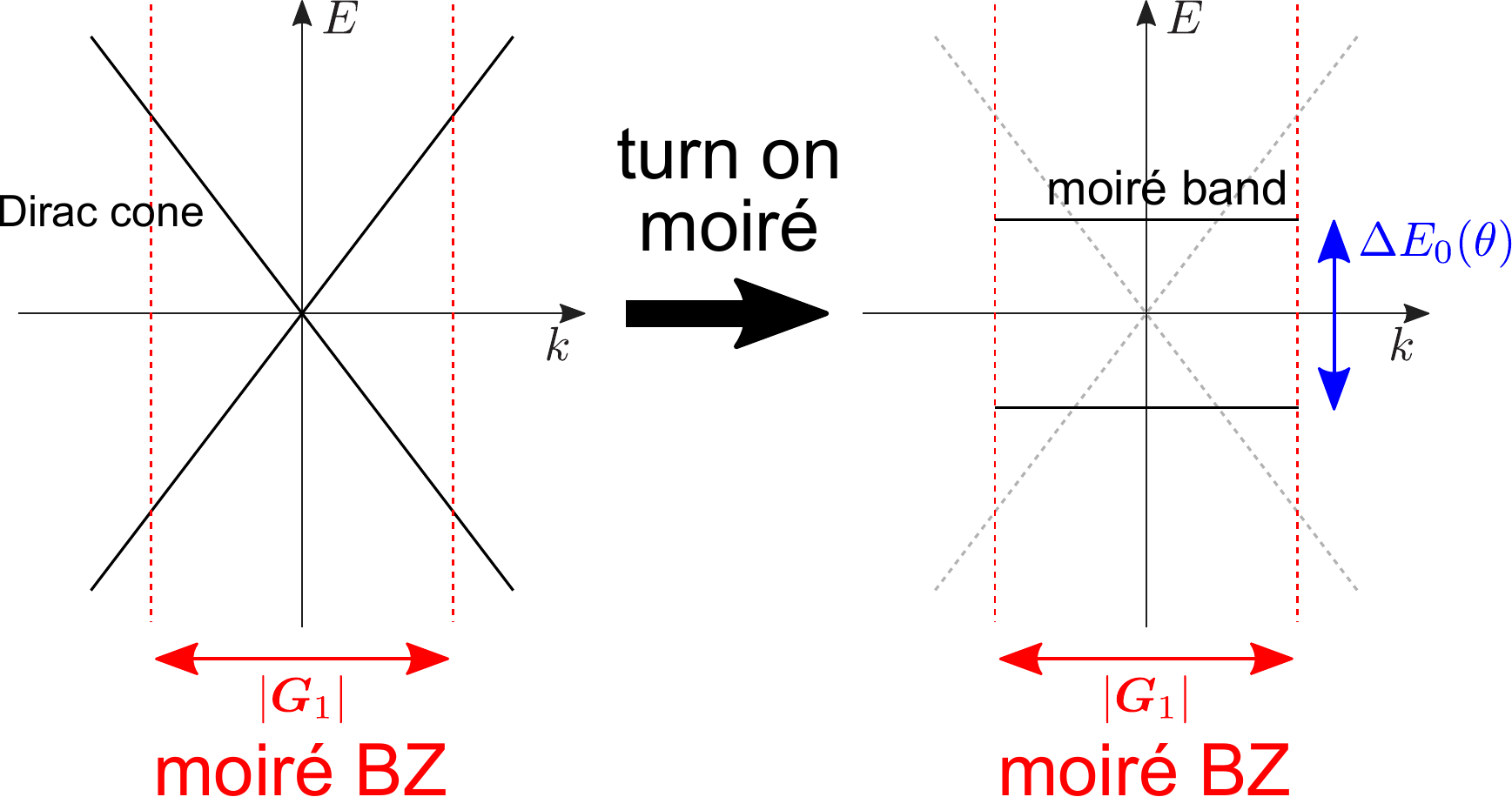}
    \caption{How to define the gap normalization factor $\Delta E_0 (\theta)$.}
    \label{fig:App_gap_normal}
\end{figure}

\section{Symmetry of a moiré system \label{sec:App_sym_of_moire}}
Generally, a twist breaks the inversion symmetry, and thus a twisted bilayer system does not have an exact inversion symmetry. The inversion symmetry breaking occurs in an atomic scale structure of the system, i. e., even if we focus on a local structure, the inversion symmetry is broken as long as the effect of twist is strictly considered. However, in the effective model with a small angle approximation, we assume that the effect of twist is negligible in the local structure and electronic states in it, and they are well approximated by an untwisted structure with a particular interlayer displacement. Therefore, if the model satisfies two conditions: (i) all sampled local structures have the inversion symmetry as untwisted structures, (ii) the layout of the local models in the moiré unit cell has the inversion symmetry, the moiré effective model recovers the inversion symmetry, although a twisted bilayer system essentially does not have the inversion symmetry in a strict sense.

In particular cases in this paper, the local model Eq.(\ref{eq:untwisted_BHZ}) has the inversion symmetry and thus condition (i) is satisfied in all cases. In the $C_3$ symmetric cases, condition (ii) is generally violated. However, (ii) can be satisfied in some special cases. For example, in the left panels in Fig.\ref{fig:C3phased1}, the layout of local models is inversion symmetric around $\bm{r}_0$ because $M_1=M_2$, and an approximate inversion symmetry is recovered in the moiré model. In the $C_2$ symmetric cases, the sampling points are inversion symmetric and thus condition (ii) is always satisfied.

\bibliography{reference}

\end{document}